\documentclass[preprint,12pt]{elsarticle}

\usepackage{graphicx}
\usepackage{color}
\usepackage[colorlinks,bookmarks=false,citecolor=blue,linkcolor=red,urlcolor=blue]{hyperref}

\usepackage{amssymb,amsbsy}
\usepackage{amsmath}

\newcommand{\bra}[1]{\ensuremath{\langle #1|}}
\newcommand{\ket}[1]{\ensuremath{|#1\rangle}}

\newcommand{\op}[1]{%
    \fontdimen12\textfont3=2pt\fontdimen12\scriptfont3=1.4pt%
    \!\null\mathop{\vphantom{#1}\smash{#1}}\limits_{\sim}\null\!}

\newcommand{\EinsOp}
           {\;\smash{\raisebox{-1.1ex}{$\!\!\stackrel{\!\mbox{1}
            \hspace{-0.4ex}\rule[0.0ex]{0.06ex}{1.60ex}}{\sim}$}}}

\newcommand{\xref}[1]{\protect\ref{#1}}
\newcommand{\figref}[1]{Fig.~\protect\ref{#1}}

\def\be{\begin{equation}}
\def\ee{\end{equation}}
\def\beq{\begin{eqnarray}}
\def\eeq{\end{eqnarray}}
\def\bc{\begin{center}}
\def\ec{\end{center}} 

\journal{J. Magn. Magn. Mater.}

\begin{document}

\begin{frontmatter}



\title{Thermal DMRG for highly frustrated quantum spin chains: a user perspective}


\author{Robin Heveling\fnref{os}}
\author{Johannes Richter\fnref{dd}}
\author{J{\"u}rgen Schnack\corref{cor1}\fnref{bi}}
\ead{jschnack@uni-bielefeld.de}
\cortext[cor1]{corresponding author}
\address[os]{Department of Physics, Osnabr{\"u}ck University,
  Barbarastr. 7, D-49076 Osnabr{\"u}ck, Germany}
\address[dd]{Institute for Theoretical Physics,
  Magdeburg University, P.O. Box 4120, D-39016 Magdeburg, Germany
  \& Max Planck Institute for Physics of Complex Systems,
        N\"{o}thnitzer Stra{\ss}e 38, D-01187 Dresden, Germany}
\address[bi]{Department of Physics, Bielefeld University, P.O. box
  100131, D-33501 Bielefeld, Germany}

\begin{abstract}
Thermal DMRG is investigated with emphasis of employability in
molecular magnetism studies. To this end magnetic observables at
finite temperature are evaluated for two one-dimensional quantum
spin
systems: a Heisenberg chain with nearest-neighbor
antiferromagnetic interaction and a frustrated sawtooth (delta)
chain. It is found that thermal DMRG indeed accurately
approximates magnetic observables for the chain as well as for
the sawtooth chain, but in the latter case only for sufficiently
high temperatures. We speculate that the reason is due to the
peculiar structure of the low-energy spectrum of the sawtooth
chain induced by frustration. 
\end{abstract}


\end{frontmatter}


\section{Introduction}
\label{sec-1}

The evaluation of magnetic properties starting from model
Hamiltonians allows one to obtain important information on
magnetic materials such as magnetic molecules or low-dimensional
magnets. A successful modeling of magnetic observables, e.g.
magnetization, heat capacity or cross sections for inelastic
neutron scattering, then leads to a concrete understanding of
magnetic exchange patterns as well as anisotropic contributions
to the Hamiltonian \cite{FuW:RMP13}.
 Nevertheless, a preferable complete and
numerically exact diagonalization of the Hamiltonian is often
spoiled by the vast dimension of the underlying Hilbert space
that grows exponentially with system size. Even if group
theoretical tools are employed to decompose the Hilbert space
into smaller orthogonal subspaces according to the available
symmetries and their irreducible
representations, the problem of exponential growth cannot be
resolved \cite{GaP:GCI93,SZP:JPI96,BCC:IC99,BCC:JCC99,Tsu:group_theory,BSS:JMMM00,Wal:PRB00,Tsu:ICA08,BBO:PRB07,ScS:PRB09,ScS:IRPC10,RiS:EPJB10}.   

Therefore, approximate methods provide a very valuable means to
assess magnetic properties of larger spin systems. Some of these
methods aim at only low-lying states, which is sufficient to
understand low-temperature spectroscopic data, others address
thermodynamic properties over a larger temperature range. Among
the approximate methods that were successfully used over the
past years are 
high-temperature series expansions (HTE) \cite{ElY:PRB94,OiB:PRB96,Oitmaa2006,TWB:DT10,SLR:PRB11,LSR:PRB14,ZRS:CAEJ16},
density-matrix renormalization group theory (DMRG \& DDMRG) \cite{Whi:PRL1992,Whi:PRB93,ExS:PRB03,Sch:RMP05,Jec:PTPS08,SCH:AoP11,UNM:PRB12},
quantum Monte Carlo (QMC) \cite{San:PRB99,SyS:PRE02,TMB:CC09,Sch:PRB16,QSC:ACIE17,BML:npjQM18},
as well as Lanczos and in particular finite-temperature Lanczos
methods (FTLM)
\cite{Lan:JRNBS50,PhysRevB.49.5065,JaP:AP00,ScW:EPJB10,HaS:EPJB14,HIL:IC16,ScT:PR17,PrK:PRB18,SSR:PRB18}.

In the present article we would like to introduce another
approximate method to the molecular magnetism community. It
extends the applicability of DMRG to non-zero temperatures
\cite{SiK:EL02,SiK:PRB02,VGC:PRL04,ZwV:PRL04,FeW:PRB05,Whi:PRL09,TMP:PRB14} and 
allows to treat very large spin systems,
in particular larger than FTLM allows, without being hampered by the
negative sign problem of QMC. The method 
-- thermal DMRG (ThDMRG) -- rests
on an approximate time evolution of matrix product states along
imaginary time from a state representing infinite temperature 
down to the temperature of interest. We will provide one example
-- a spin chain -- where this approximation yields very accurate
results and another one -- a sawtooth chain-- where the approximation
fails for low temperatures.

The article starts with a short reminder of the thermal DMRG in
Sec.~\ref{sec-2}, presents our results in Sec.~\ref{sec-3}, before 
it summarizes our main points in  Sec.~\ref{sec-4}.

\section{Thermal DMRG}
\label{sec-2}

The idea of finite-temperature, i.e. thermal DMRG (ThDMRG) is
published in several references
\cite{SiK:EL02,SiK:PRB02,VGC:PRL04,ZwV:PRL04,FeW:PRB05,Whi:PRL09,TMP:PRB14}.
Here we would like to provide a very short description of the
idea.

A density matrix at temperature $T$ and external field $B$ can
be thought of as the time-evolved density matrix starting from a
density matrix at infinite temperature down to the desired
temperature $T$ by means of imaginary time evolution with the
appropriate Hamiltonian $\op{H}(B)$ of the system
\begin{eqnarray}
\label{E-2-1}
\op{\rho}(\beta,B)
&=&
\exp\left[
-i \op{H}(B) \tau/\hbar
\right]
\op{\rho}(\beta=0)
\exp\left[
i \op{H}(B) \tau/\hbar
\right]
\\
\text{with}
&&
\tau=-i\beta\hbar/2\ , \quad \beta = \frac{1}{k_B T}
\ .
\end{eqnarray}
The inverse temperature $\beta$ plays the role of the imaginary
time. The density matrix at infinite temperature is proportional
to the unit operator.

This approach is chosen since one can then employ time-dependent
DMRG for the approximate time evolution of a quantum state
describing the large spin system. Technically, the
time evolution of the density matrix is replaced by the time
evolution of a state vector which is first constructed by
purifying the density matrix. Purification turns a density
matrix into a ket state belonging to a
larger Hilbert space in a way, that a trace over the additional
(auxiliary) degrees of freedom returns the desired density
matrix \cite{KKM:PRA06,NoA:PRA16,Bar:PRB16}. 

\begin{figure}[ht!]
\centering 
\includegraphics[width=0.90\textwidth]{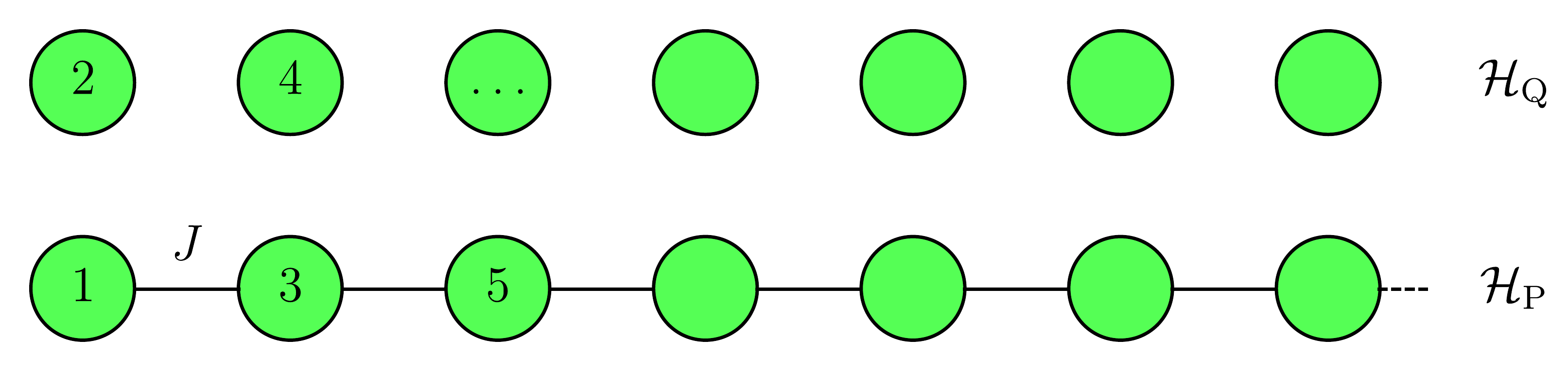}
\caption{Visualization of the enlarged system for a physical
  spin chain: Green circles connected by black lines mark the
  spins of the chain, whereas the green circles on top denote
  ancilla spins. They do not interact with the physical
  system. \label{thdmrg-f-1}}
\end{figure}

In particular one performs the following steps:
\begin{itemize}
\item Set up the initial state $\ket{\Psi_0}$ that corresponds to infinite
  temperature. This state is a matrix product state in a Hilbert
  space consisting of the Hilbert space ${\mathcal H}_P$ of the
  physical system one wants to model as well as the Hilbert
  space ${\mathcal H}_Q$ of the unphysical system of
  auxiliary degrees of freedom called \emph{ancillas},
  compare \figref{thdmrg-f-1}.
  There is a certain freedom in constructing $\ket{\Psi_0}$
  \cite{Sch:AP11,NoA:PRA16}. In our
  implementation we employ the entangler Hamiltonian
  \cite{NoA:PRA16} to generate $\ket{\Psi_0}$.

\item Evolve the state in imaginary time by means of
  time-dependent DMRG
  \cite{cazalilla,schmitti,feiguin,daley}
\begin{equation}
\label{Img}
\ket{\Psi_\beta} \equiv e^{-\beta \op{H}(B)/2}\ket{\Psi_0}
\ .
\end{equation}
 This is the numerically demanding part since the basis, which was
 optimized for the 
  state at $\beta=0$, is not well-enough suited for later times
  and therefore must be adapted by targeting evolved states at
  certain times. For this purpose we project the time evolution
  operator onto a Krylov subspace \cite{MMN:ACP05}. Other time
  evolution algorithms 
  such as Runge-Kutta or Chebyshev are equally feasible. A
  Suzuki-Trotter approach works best for nearest-neighbor
  interactions. Since additional long-range interactions are
  introduced by 
  purifying the state, this method may not be optimal
  here. Grouping spins together to recover nearest-neighbor
  interaction is possible but quickly becomes tedious. 

\item Finally, calculate the thermal expectation value of a
  physical observable (not acting on the auxiliary degrees of
  freedom)
\begin{equation}
\langle \,\op{O}\,\rangle_{\beta} =
\dfrac{\bra{\Psi_\beta}\op{O}\ket{\Psi_\beta}}{
  \bra{\Psi_{\beta}} \Psi_{\beta} \,\rangle}
\ ,\quad
\op{\rho}(\beta)= \dfrac{Z(0)}{Z(\beta)}\,\text{Tr}_{\text{Q}} \ket{\Psi_\beta} \bra{\Psi_\beta}
\ ,
\end{equation}
with
\begin{equation}
\EinsOp=Z(0)\op{\rho}(0)\ ,\quad
Z(\beta) = Z(0)\, \bra{\Psi_{\beta}} \, \Psi_{\beta}\,\rangle
\ .
\end{equation}
$\text{Tr}_{\text{Q}}$ denotes the partial trace in ${\mathcal H}_Q$.

\item The accuracy of this method depends on a few parameters,
  among which the number $m$ of kept basis states plays a
  major role. We use $m$ values of up to 60, and work with fixed
  $m$. Alternatively, one could employ the discarded weight of
  DMRG as a measure of quality; this would lead to varying $m$
  values in the course of calculations to keep the discarded
  weight constant. An important part of 
  basis improvement are the so-called sweeps. We run one full
  sweep in-between time steps, which is usually enough to obtain
  a sufficiently adapted basis. One may choose to perform more
  sweeps to increase accuracy \cite{FeW:PRB05B}. The time step $dt$ for the
numerical time evolution is chosen as $dt=0.1$ (in natural
units). Smaller $dt$ did not yield better results; on the
contrary, one has to trade off between more accurate shorter
time steps and the total time duration one can achieve.

\end{itemize}
In this outlined way one evaluates observables for decreasing
temperatures (increasing $\beta$). The calculation has to be
repeated for every field value $B$ one is interested in.

\section{Numerical examples}
\label{sec-3}

\begin{figure}[ht!]
\centering 
\includegraphics[width=0.90\textwidth]{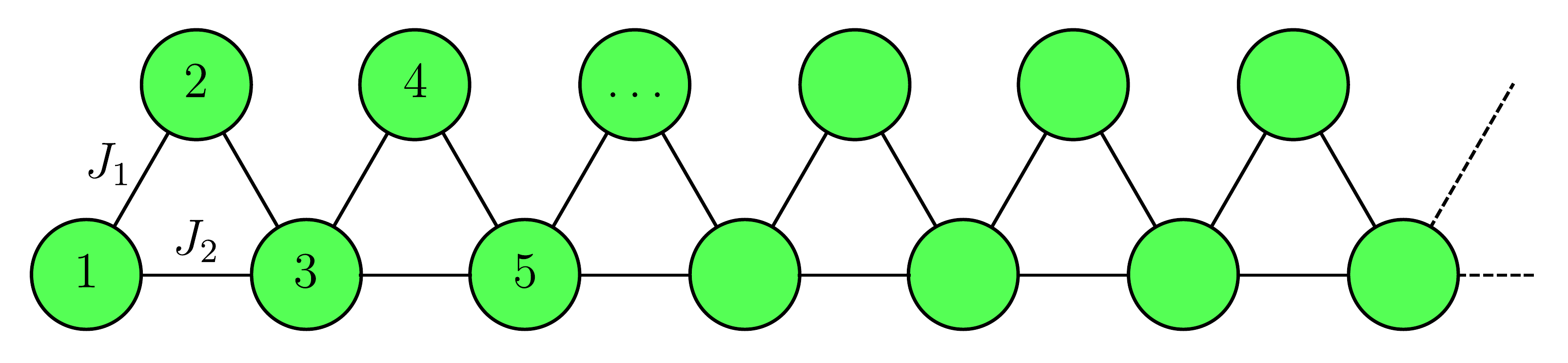}
\caption{Visualization of the Hamiltonian for the sawtooth
  chain: Green circles mark the sites, solid lines the
  interactions of the physical system with Hilbert space
  ${\mathcal H}_P$. The sketched situation corresponsd to OBC
  with an odd number of spins. OBC with an even number of spins
  result in a dangling $J_1$ bond with the terminal
  spin. \label{thdmrg-f-z}}  
\end{figure}

The following examples are evaluated for the respective
Heisenberg Hamiltonian
\begin{eqnarray}
\label{E-I-1}
\op{H}
&=&
-2\;
\sum_{i<j}\;
J_{ij}
\op{\vec{s}}_i \cdot \op{\vec{s}}_{j}
+
g \mu_B B
\sum_{i}\;
\op{{s}}_i^z
\ ,
\end{eqnarray}
where $\op{\vec{s}}$ denotes the quantum spin operators. In the
following examples the single-spin quantum number is always
$s=1/2$. 
For the linear chain with antiferromagnetic nearest-neigbor
exchange interaction $J_{ij}=J$ for adjacent spins on the chain;
for the sawtooth chain the spins are connected by $J_1$ and
$J_2$ as depicted in \figref{thdmrg-f-z}.

\subsection{Spin chain}
\label{sec-3-1}

\begin{figure}[ht!]
\centering
\includegraphics*[clip,width=0.5\columnwidth]{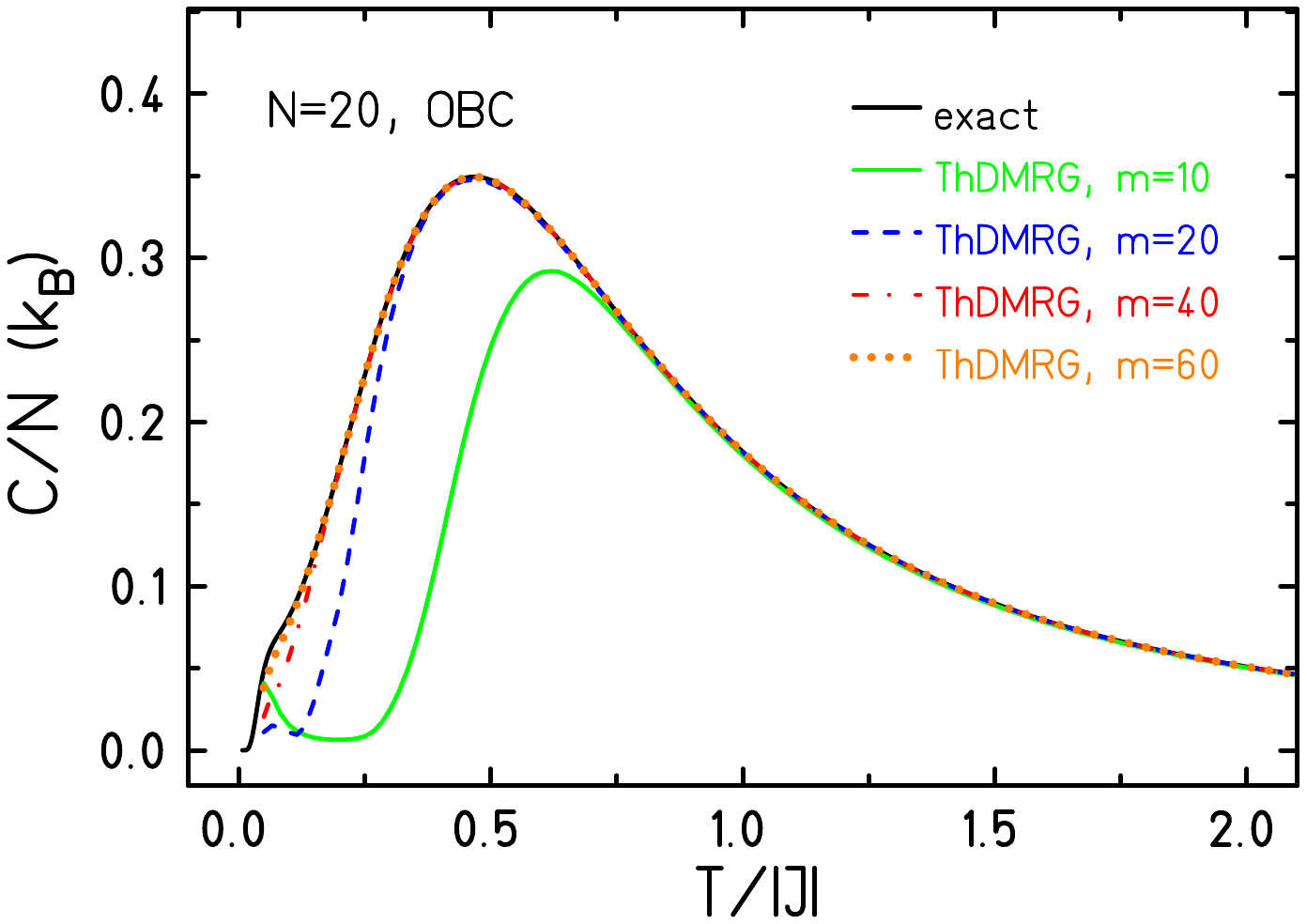}

\includegraphics*[clip,width=0.5\columnwidth]{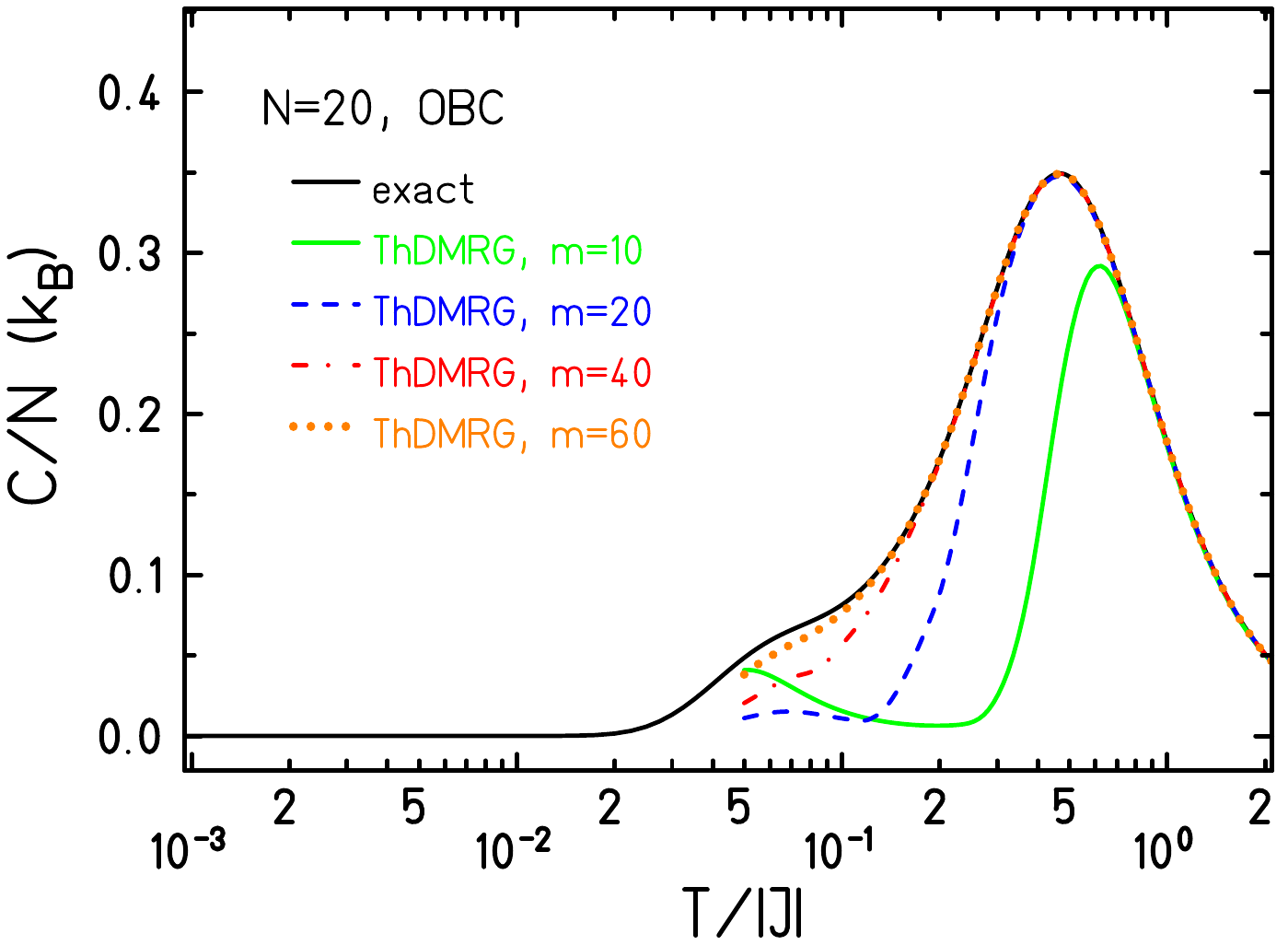}

\includegraphics*[clip,width=0.5\columnwidth]{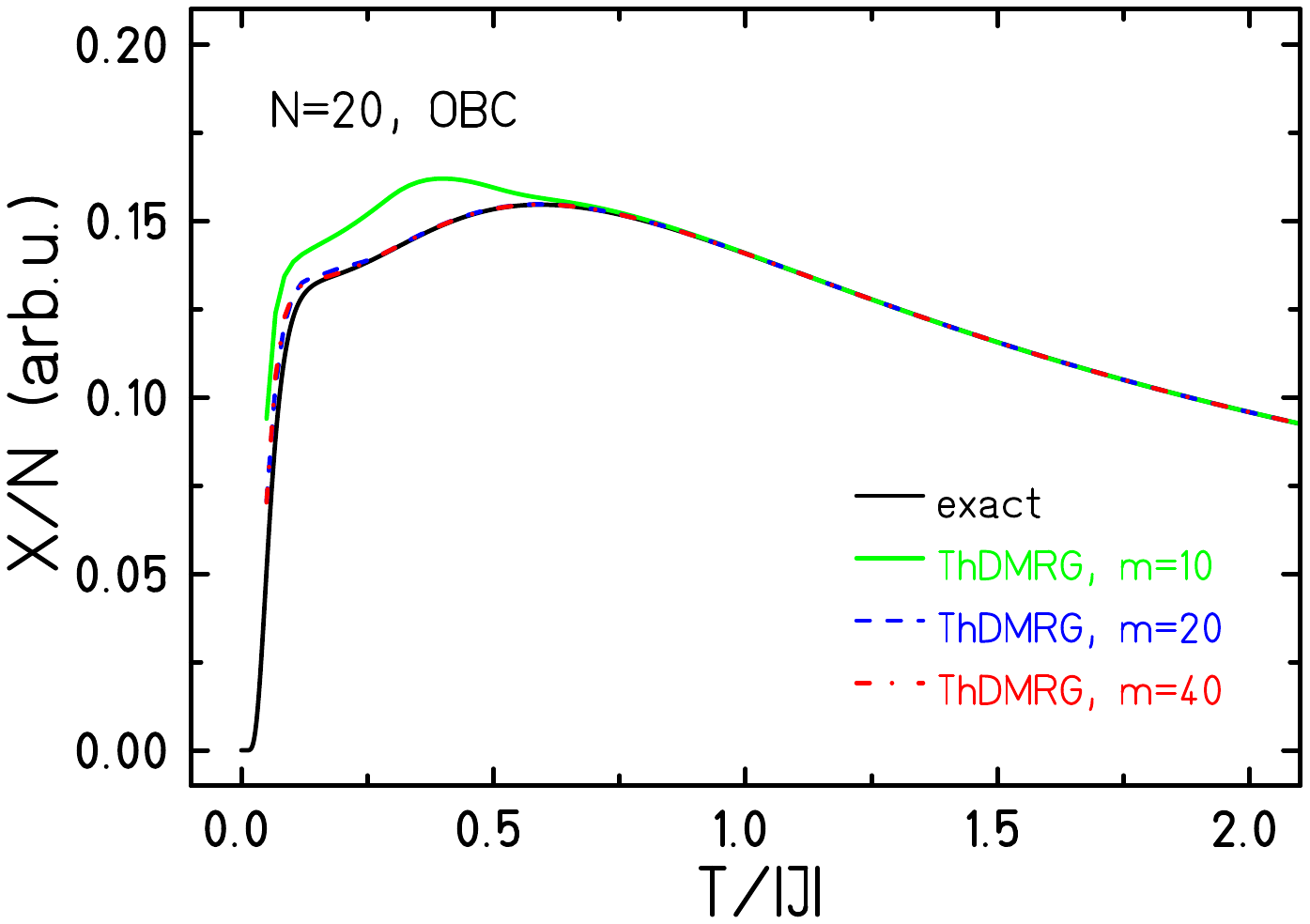}
\caption{Spin-$1/2$ chain of length $N=20$:  The zero-field specific heat
  capacity (top \& center) and the specific low-field susceptibility 
  (bottom) are plotted vs temperature for $m=10$ (green),
  $m=20$ (blue), $m=40$ (red) and $m=60$ (orange). The black
  solid curves represent exact diagonalization data.}   
\label{thdmrg-f-d}
\end{figure}

As our first example we would like to discuss the evaluation of
magnetic observables for a linear spin chain of $N=20$ spins
$s=1/2$ with antiferromagnetic nearest-neighbor interaction.
This example, which was already investigated by Feiguin and White
\cite{FeW:PRB05}, serves to show that the method is indeed capable of
reproducing exact results with rather high
accuracy. Figure~\xref{thdmrg-f-d} shows the specific heat
capacity (top \& center) and the specific zero-field susceptibility
(bottom) plotted vs temperature. The values $m=10$ (green), $m=20$ (blue),
$m=40$ (red), and $m=60$ (orange) have been used in the
approximation; black solid curves represent the respective
observables obtained from exact diagonalization \cite{ScS:IRPC10,HeS:19}.
 The zero-field
specific heat is evaluated using the variance of the
Hamiltonian, whereas the susceptibility is evaluated as
difference quotient, i.e.,
\begin{equation}
\label{magapp}
\chi(T) = \dfrac{\partial M(T,B)}{\partial B}\bigg{|}_{B=0}
\approx \dfrac{M(T,0.1) - M(T,0)}{0.1 - 0}=
\dfrac{M(T,0.1)}{0.1}
\ . 
\end{equation}
Interestingly, this yields better results than calculating the
susceptibility as a variance of the magnetization. As can be
seen in the panels of \figref{thdmrg-f-d} the results
improve as $m$ increases. For the susceptibility the
approximation lies almost on top of the ED curve already for $m=20$ and
$m=40$, whereas the low-temperature shoulder of the specific heat
needs at least $m=60$ to be reasonably well approximated,
compare \figref{thdmrg-f-d} (center).

\subsection{Sawtooth chain}
\label{sec-3-2}

The situation changes drastically when moving to a sawtooth (delta)
chain system with ferromagnetic interaction $J_1$ between
adjacent apical and basal spins and
antiferromagnetic interaction $J_2$ between adjacent basal
spins, compare \figref{thdmrg-f-z}.   
We selected this system on purpose since it
exhibits a quantum phase transition for $\alpha=|J_2/J_1|=1/2$
\cite{KDN:PRB14,DmK:PRB15,BML:npjQM18}. 
At the transition point there is a flat one-magnon 
excitation band that leads to extraordinary features in the
low-energy spectrum \cite{KDN:PRB14}.
In particular, 50~\% of the total available entropy is contained in the
massively degenerate ground state. Moreover, even for finite sawtooth chains  
the exication gap is extremely small and the excitations above the massively 
degenerate ground-state mani\-fold are also highly degenerate. As a result, 
low-temperature physics is unconventional \cite{KDN:PRB14}.
In the vicinity of the
critical point the degeneracy splits up and leads to an
additional well-pronounced
low-temperature maximum of the heat capacity. The position of
the maximum corresponds to an additional low-temperature scale
$T_{\text{low}}$. One would expect that the imaginary time
evolution of the DMRG matrix product state runs into problems
when this temperature scale is reached from above and thus 
the state $\ket{\Psi_\beta}$ enters a part of Hilbert space with
an enormous spectral density and entropy contribution.

\begin{figure}[ht!]
\centering
\includegraphics*[clip,width=0.5\columnwidth]{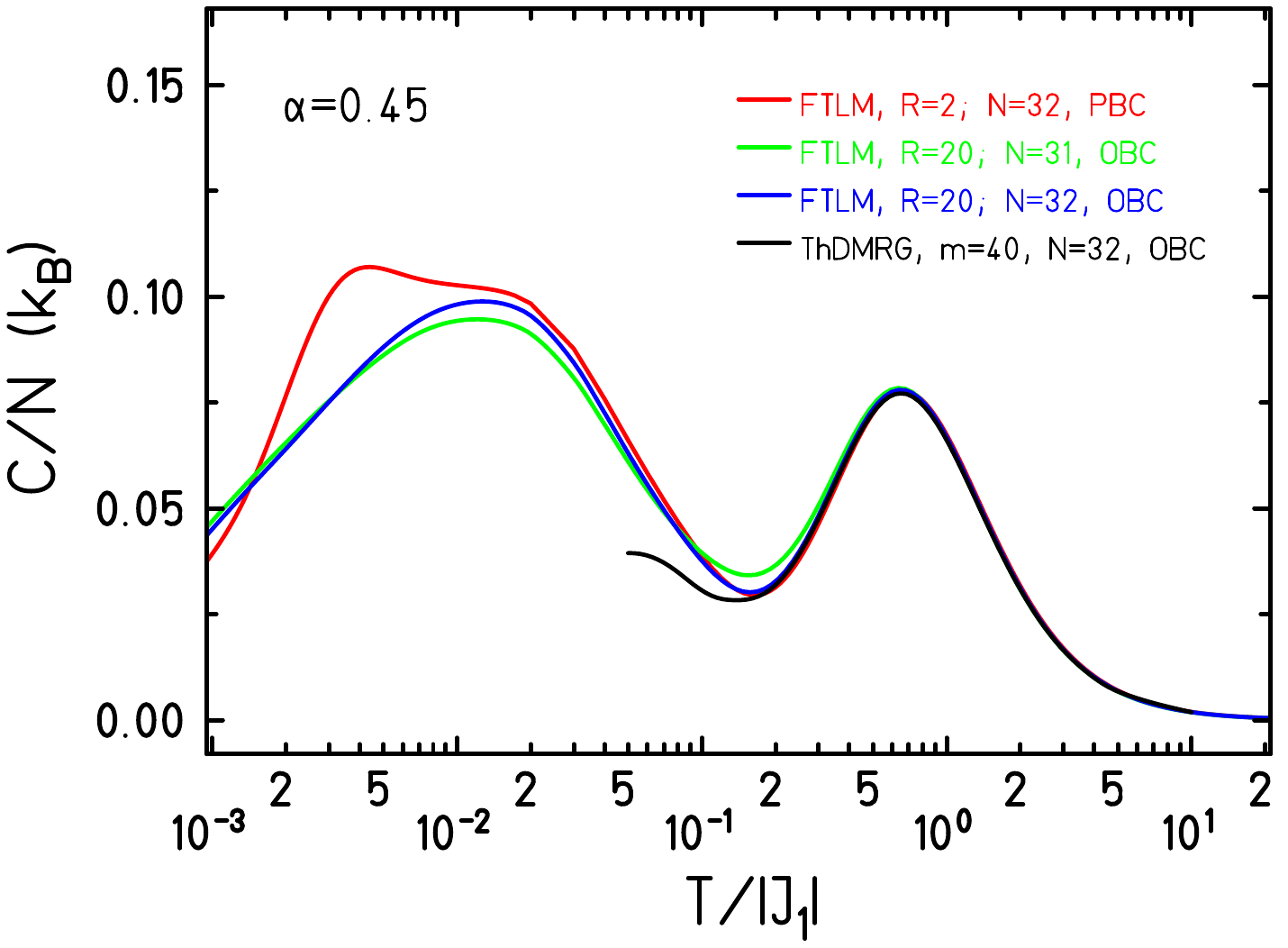}

\includegraphics*[clip,width=0.5\columnwidth]{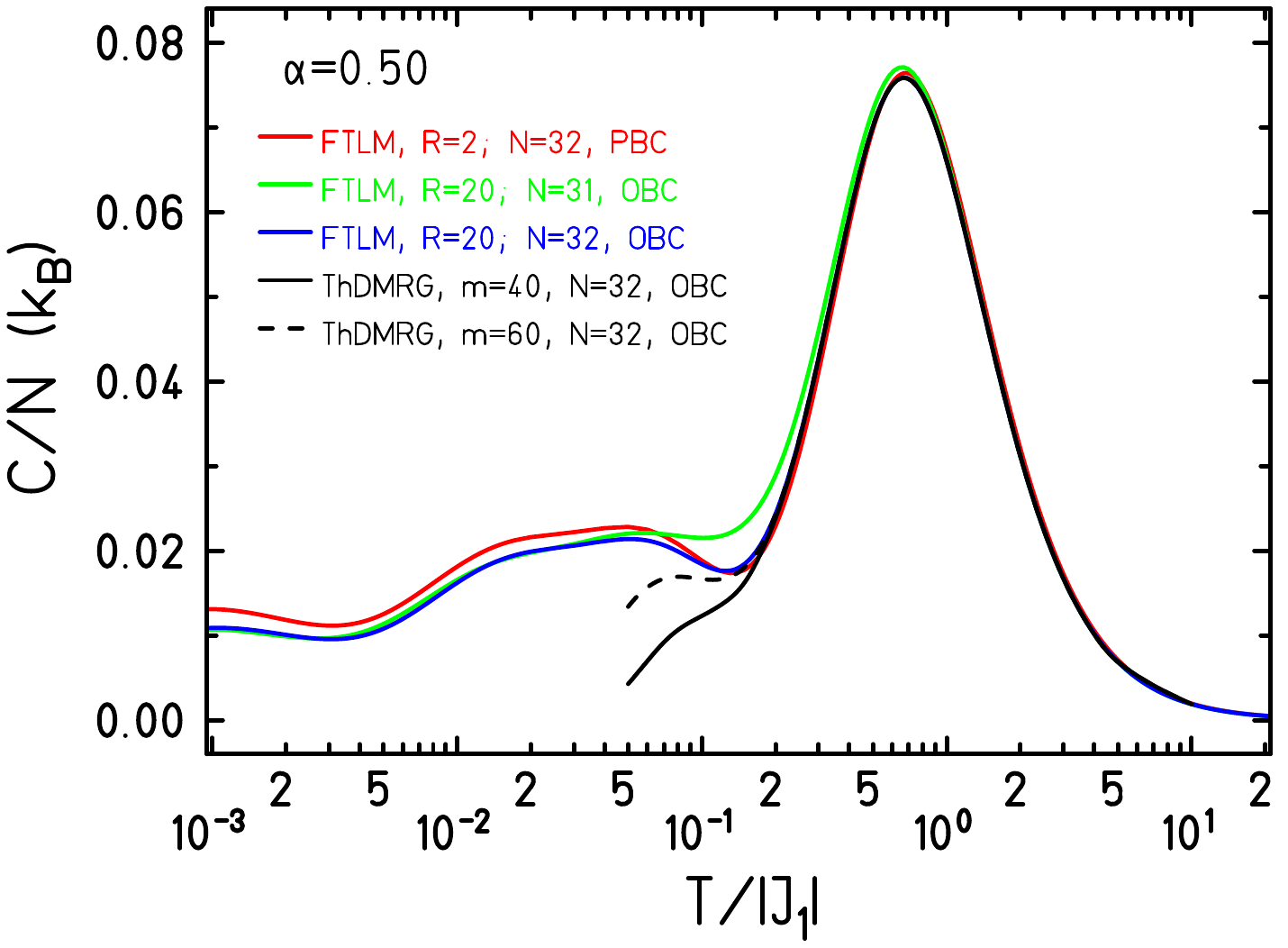}

\includegraphics*[clip,width=0.5\columnwidth]{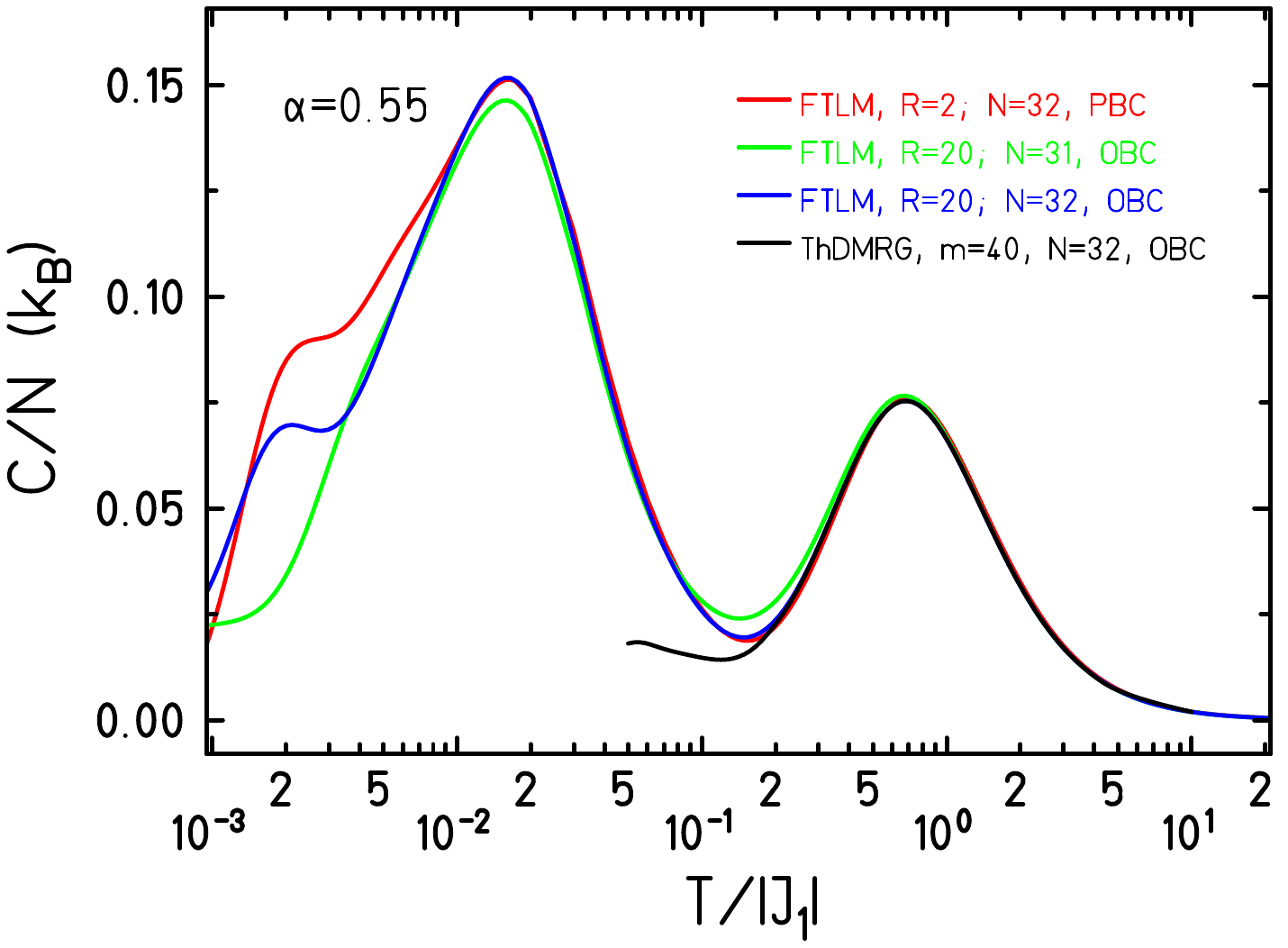}
\caption{Specific heat of the sawtooth chain for $\alpha=0.45$
  (top), $\alpha=0.50$ (center) and $\alpha=0.55$ 
(bottom): thermal DMRG (black curves), various related FTLM
  results (colored curves). See text for detailed explanations. } 
\label{thdmrg-f-a}
\end{figure}

Figure~\xref{thdmrg-f-a} displays the results for a sawtooth
chain with $N=32$ and open boundary conditions (OBC) as
black solid curves for $\alpha=0.45$ (top), $\alpha=0.50$
(center) and $\alpha=0.55$ 
(bottom); $m=40$ was used for the calculations. This result is
compared to FTLM approximations which 
are virtually exact as discussed elsewhere \cite{SSR:PRB18}. 
In the case of periodic boundary conditions (PBC) translational
symmetry was employed by means of the 
freely available program \verb§spinpack§ \cite{spin:256}
(Therefore, much smaller $R$ suffice.).
The blue curves display FTLM results that correspond to the
situations investigated by ThDMRG, the red and green curves show
FTLM results of closely related cases of $N=32$ (PBC) as well as 
$N=31$ (OBC, no dangling bond, see \figref{thdmrg-f-z}),
respectively. As one can 
visually verify, the approximation by means of thermal DMRG
becomes untrustworthy when the temperature approaches the
low-energy scale $T_{\text{low}}$, i.e., the realm of the
low-temperature maximum. The reason is that the low-energy part
of the spectrum and thus of 
Hilbert space can (effectively as a whole)
be easily represented by a matrix
product state (of smaller $m$) as long as the state
$\ket{\Psi_\beta}$ resides at higher temperatures. But at
temperatures of the order of $T_{\text{low}}$ and below, the matrix product state
would need to resolve the spectral decomposition at this scale,
which becomes virtually impossible in view of the vast dimension that
is exponential in system size. This phenomenon worsens the
problem of time-dependent DMRG to become unreliable after a
characteristic \emph{runaway time}
\cite{GKS:PRE05,ScW:AIP06}. Independent of this time 
scale, $T_{\text{low}}$ introduces a system specific scale where
ThDMRG breaks down. An increase of the block dimension from
$m=40$ to $m=60$ improves the situation somewhat, compare center
of \figref {thdmrg-f-a}, but only for a limited temperature
range. In order to maintain a good accuracy an exponential
increase of block dimension would be necessary with time
\cite{GKS:PRE05,ScW:AIP06}. 

Despite the failure of thermal DMRG at very low temperatures one
must acknowledge that it successfully approximates the thermal
behavior above $T_{\text{low}}$. In particular, the
Schottky peak around $T\sim|J_1|\sim|J_2|$ is very well
reproduced. 

\begin{figure}[ht!]
\centering
\includegraphics*[clip,width=0.45\columnwidth]{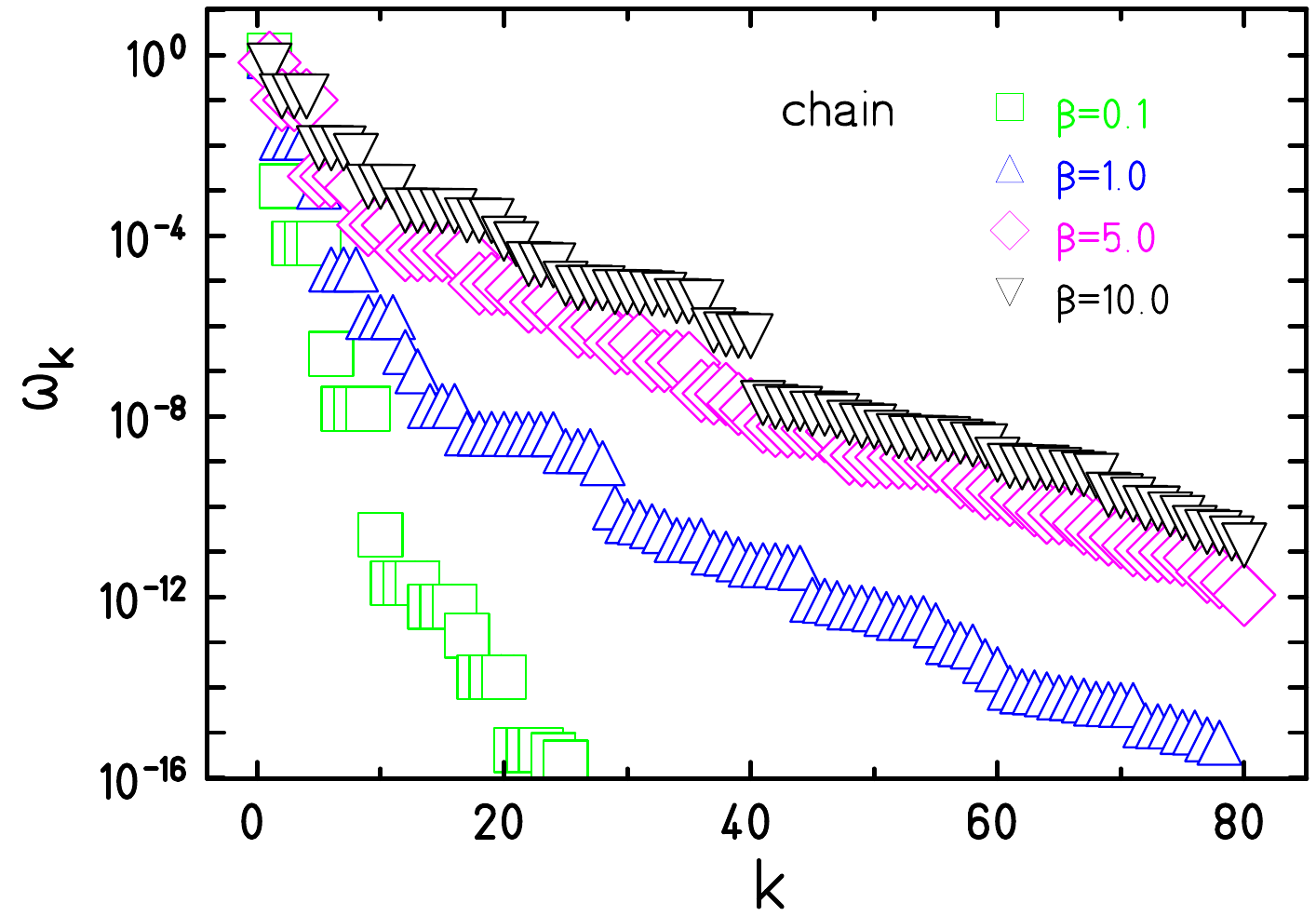}
\quad
\includegraphics*[clip,width=0.45\columnwidth]{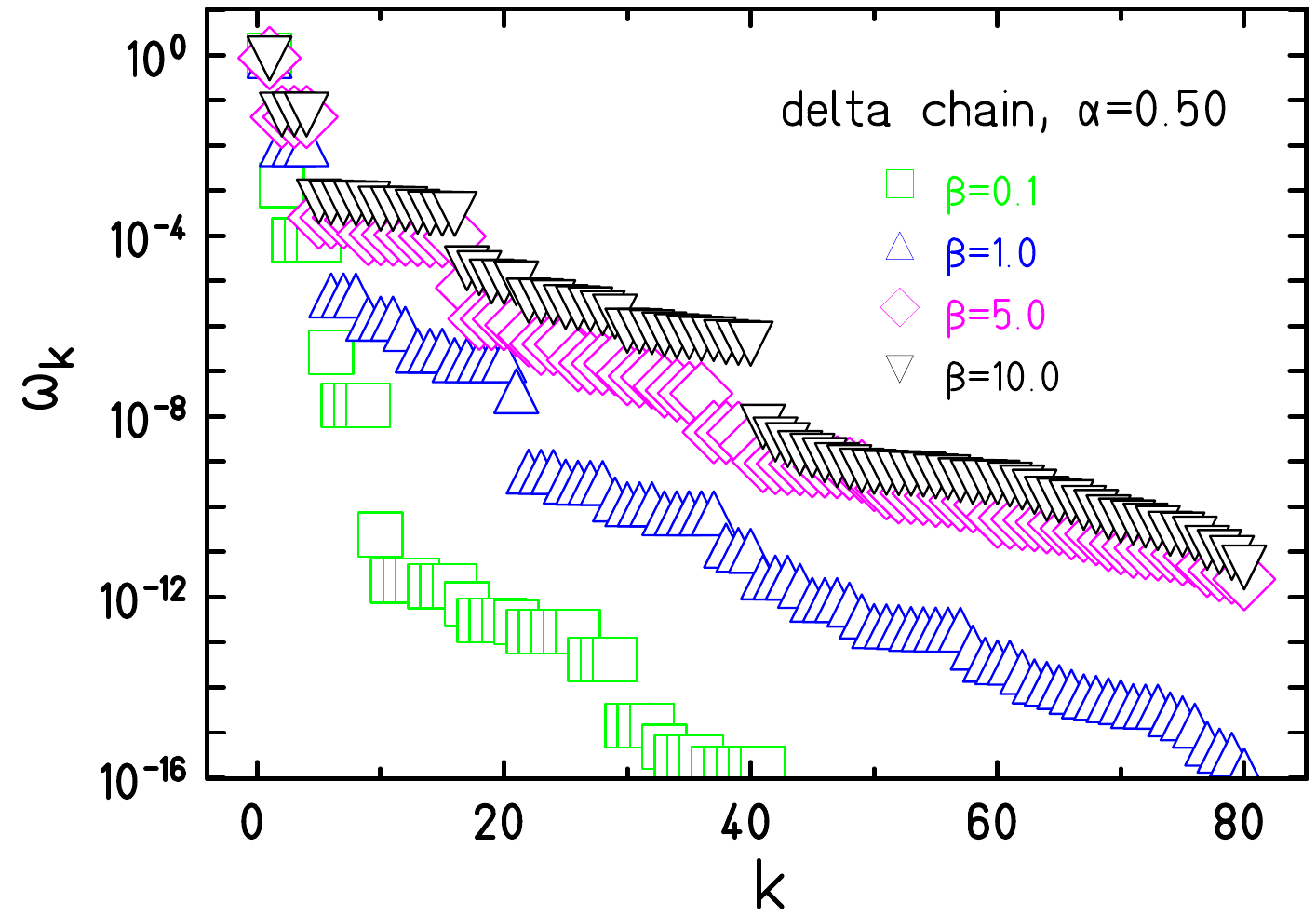}
\caption{Entanglement spectra of the reduced density matrices of
  block $A$ for a chain (l.h.s.) and a sawtooth (delta) chain with
  $\alpha=0.5$ (r.h.s.) of length $L=32$: The number of states
  kept is $m=40$, OBC are applied. The imaginary time is
  provided in the figures.} 
\label{thdmrg-f-b}
\end{figure}

In order to go a bit into details, we would like to discuss the
behavior of the entanglement spectrum with imaginary time. This
serves as a measure of how good the matrix product state can
approximate the target state, which could for instance be the
ground state or $\ket{\Psi_\beta}$. A spectrum that does not
fall off rapidly, signals that a subsystem, i.e. block $A$, is
strongly entangled with the environment, i.e. block $B$, and thus
-- for the chosen value of $m$ -- the matrix product state is
not a good approximation of the target state. 

Figure~\xref{thdmrg-f-b} shows on the l.h.s. the entanglement
spectra of the reduced density matrices of block $A$ (half the
system size) for a chain and on the r.h.s. for a sawtooth
(delta) chain with 
$\alpha=0.5$, both of length $L=32$. One notices that for both systems the
spectra fall off slower and slower with increasing imaginary time,
i.e. inverse temperature. This means that the entanglement with
the environment grows and the matrix product state of fixed
block dimension $m=40$ is less and less capable to represent the
thermal state at the respective $\beta$. This effect is somewhat
more pronounced for the sawtooth chain, but not as dramatically
as the poor approximation of the low-temperature observables
would suggest. This means that the entanglement spectrum alone
is not a sufficient quantifier of the achieved quality of
calculation.

\section{Summary}
\label{sec-4}

In this article the thermal DMRG was applied to two archetypical
spin systems -- the linear antiferromagnetic chain with
nearest-neighbor exchange interaction as well as the frustrated
sawtooth (delta) chain close to the quantum critical point. Our
intention was an investigation of the achievable accuracy in
view of a later use of the method in molecular magnetism. The general
conclusion seems to be that non-trivial low-temperature features
are hard to resolve and approximate properly, since such
features stem from intricate low-energy spectral properties,
which would require a good representation by the matrix product
state at this temperature. The latter becomes increasingly
unlikely since both, the intricate low-energy spectrum as well
as the unavoidable \emph{runaway time} of DMRG impede a proper
representation of the thermal state of the system in terms of a
product state of low-dimensional matrices.

Nevertheless, the accuracy for thermal magnetic observables is
very good above those scales. And, most importantly, the major
advantage to be able to treat very large and possibly frustrated
spin systems cannot be accomplished by any other approximate
method.

\section*{Acknowledgment}

This work was supported by the Deutsche Forschungsgemeinschaft DFG
(314331397 (SCHN 615/23-1); 355031190 (FOR~2692); 397300368 (SCHN~615/25-1)). 
The authors thank Andreas Kl{\"umper} and Salvatore Manmana for
helpful discussions.



\end{document}